\documentclass[12pt]{article}
\newcommand{\eq}{\begin{equation}}
\newcommand{\feq}{\end{equation}}
\newcommand{\eqn}{\begin{eqnarray}}
\newcommand{\feqn}{\end{eqnarray}}
\newcommand{\arr}{\begin{eqnarray*}}
\newcommand{\farr}{\end{eqnarray*}}
\newcommand{\beq}{\begin{equation}}
\newcommand{\eeq}{\end{equation}}
\newcommand{\bea}{\begin{eqnarray}}
\newcommand{\eea}{\end{eqnarray}}

\def\AP{ Ann. Phys. }

\def\beq{\begin{equation}}
\def\eeq{\end{equation}}
\def\bea{\begin{eqnarray}}
\def\eea{\end{eqnarray}}
\def\bc{\begin{displaymath}}
\def\ec{\end{displaymath}}
\def\lb{\label}
\def\tlambda{\tilde\lambda}
\def\tgtt{\tilde\ga_{tt}}
\def\tgxx{\tilde\ga_{rr}}
\def\tgff{\tilde\ga_{\f\f}}
\def\gtt{\ga_{tt}}
\def\gxx{\ga_{rr}}
\def\gff{\ga_{\f\f}}
\def\ha{{1\over2}}
\def\gr{\sqrt{-g}}
\def\Ht{{\cal H}}
\def\Hx{{\cal H}_r}
\def\Pe{\Pi_\Phi}
\def\Ps{\Pi_\sigma}
\def\xo{{\xi^\perp}}
\def\xp{{\xi^\parallel}}
\def\oo{{\chi^\perp}}
\def\op{{\chi^\parallel}}
\def\ds{ds^2=}
\def\mo{{-1}}
\def\eo{\Phi_0}

\def\al{\alpha}
\def\be{\beta}
\def\ga{\gamma}
\def\de{\delta}
\def\ep{\varepsilon}
\def\f{\phi}

\def\si{\sigma}\def\s{\sigma}

\def\om{\omega}

\def\r{\rho}

\def\La{\Lambda}

\def\lb{\label}
\def\nn{\nonumber}
\begin{document}
\begin{titlepage}
\begin{flushright}
INFNCA-TH0104 \\
MIT-CTP-3134 \\
\end{flushright}
\vspace{.3cm}
\begin{center}
\renewcommand{\thefootnote}{\fnsymbol{footnote}}
{\Large \bf Conformal Dynamics of 0-Branes}
\vfill
{\large \bf {M.~Cadoni$^{1,a}$\footnote{email: cadoni@ca.infn.it}, 
P.~Carta$^{1,a}$\footnote{email: carta@ca.infn.it},
M. Cavagli\`a$^{2,b}$\footnote{email: cavaglia@mitlns.mit.edu}
and S.~Mignemi$^{3,a}$\footnote{email: mignemi@ca.infn.it}}}\\
\renewcommand{\thefootnote}{\arabic{footnote}}
\setcounter{footnote}{0}
\vfill
{\small
$^1$ Universit\`a di Cagliari, Dipartimento di Fisica,\\
Cittadella Universitaria, 09042 Monserrato, Italy\\
\vspace*{0.4cm}
$^{2}$ Center for Theoretical Physics, Massachusetts Institute of Technology,
77 Massachusetts Avenue, Cambridge MA 02139-4307, USA\\
\vspace*{0.4cm}
$^3$ Universit\`a di Cagliari, Dipartimento di Matematica,\\
Viale Merello 92, 09123 Cagliari, Italy\\
\vspace*{0.4cm}
$^a$ INFN, Sezione di Cagliari\\
\vspace*{0.4cm}
$^b$ INFN, Sede di Presidenza, Roma}
\end{center}
\vfill
\centerline{\bf Abstract}
\vfill
We investigate the dynamics of dilatonic D-dimensional 0-branes in the
near-horizon regime. The theory is given in a twofold form: two-dimensional
dilaton gravity and nonlinear sigma model. Using asymptotic symmetries, duality
relations, and sigma model techniques we find that the theory has three
conformal points which correspond to (a) the asymptotic (Anti-de Sitter) region
of the two-dimensional spacetime, (b) the horizon of the black hole, and (c)
the infinite limit of the coupling parameter. We show that the conformal
symmetry is perturbatively preserved at one-loop, identify the corresponding
conformal field theories, and calculate the associated central charges.
Finally, we use the conformal field theories to explain the thermodynamical
properties of the two-dimensional black holes.
\vfill
\end{titlepage}
\section{Introduction}
Dilatonic 0-branes are solutions of D-dimensional supergravity coupled to
$U(1)$ gauge fields that describe effective low-energy approximations to the
D0-brane solutions of string theory. Their investigation is relevant to
understanding the Anti-de Sitter/Conformal Field Theory (AdS/CFT)
correspondence \cite{adscft} in two spacetime dimensions \cite{ads,
Cadoni:1999ja, Maldacena:1999uz,Cadoni:2001fq}. Though the AdS$_2$/CFT$_1$
correspondence is quite well-known in the dilaton gravity context
\cite{Cadoni:1999ja,Cadoni:2001fq}, little is known in the more general
framework of string theory.

In a recent paper \cite{Youm:2000vn} Youm has shown that in the dual-frame
near-horizon regime D-dimensional dilatonic 0-branes can be described by an
effective two-dimensional dilaton gravity model with non-constant dilaton and
asymptotically AdS$_2$ black hole solutions. The SL(2,R) isometry group  of
AdS$_2$ is thus broken, a feature which has prevented any attempt to using the
asymptotic symmetries of AdS$_2$ to generate an infinite-dimensional conformal
symmetry associated with the dynamics of the 0-brane. However, previous
investigations \cite{Cadoni:2000ah,navarro,CV,CCV} of the Jackiw-Teitelboim
model \cite{JT}  (which describes the near-horizon behavior of a specific
0-brane) have shown that the breaking of conformal symmetry due to a nontrivial
dilaton can actually be controlled and is essential to understanding features
of the CFT such as the existence of a nonvanishing central charge in the
Virasoro algebra \cite{Cadoni:2000ah}. Thus, applying similar arguments we can
investigate the existence of conformal symmetries for a general 0-brane.

In this paper we show that the asymptotic symmetries of the near-horizon
0-brane solutions are generated by a Virasoro algebra. Using a canonical
realization of the asymptotic symmetries we calculate the central charge of the
algebra and give an explicit realization of the conformal symmetry in terms of
the fields that describe deformations of the boundary of AdS$_2$. For a
particular range of the coupling parameter $a$ we identify the one-dimensional
conformal mechanics that lives on the boundary of AdS$_2$ and realizes the
conformal symmetry. In the limit $a\to\infty$ the dilaton gravity model is
shown to be equivalent to a free CFT. Thanks to a previous result by Carlip
\cite{Carlip:1999cy} we also argue that the horizon of the two-dimensional
black hole defines a CFT with well-defined central charge. In the sigma model
formulation the existence of these three conformal points is recovered, at the
classical level, by  relating the asymptotic symmetries of the gravitational
theory to the conformal symmetries of the sigma model and by implementing the
duality symmetries of the  theory. The calculation of one-loop beta functions
shows that the conformal symmetry is perturbatively preserved. Finally, the CFT
results are used to discuss the thermodynamical behavior of the two-dimensional
black holes.

The structure of the paper is the following. In Sect.\ 2 we review the
two-dimensional dilaton gravity model that describes the near-horizon regime of
D-dimensional dilatonic 0-branes. We consider the (asymptotically AdS$_2$)
black hole solutions and discuss different limiting cases in the moduli space
of the theory. In Sect.\ 3 we investigate the asymptotic symmetry group (ASG)
of the solutions. We show that the ASG is generated by a Virasoro algebra and
calculate the central charge. We discuss the dynamics induced on the AdS$_{2}$
boundary by the bulk gravity theory. In Sect.\ 4 and 5 we identify the CFT that
lives on the horizon of the black hole and use the sigma model formulation to
describe the relation between the conformal symmetries of the sigma model and
the asymptotic symmetry group of dilaton gravity. We also show that the
weak-coupling region can be described by a free CFT. Sect.\ 6 deals with the
duality symmetries of the theory. The sigma model approach is used to prove
that the horizon of the black hole defines a CFT. In Sect.\ 7 we calculate the
one-loop beta functions of the three CFTs and show that the conformal symmetry
is preserved at one-loop at the (classical) conformal points. In Sect.\ 8 we
discuss the thermodynamical properties of the two-dimensional black holes.
Finally, we state our conclusions in Sect.\ 9.
\section{Effective theory of dilatonic 0-branes}
In the Einstein frame the bosonic part of the supergravity action that
describes dilatonic 0-branes solutions in D dimensions is
\eq\lb{suact}
S = \frac{1}{2\kappa^{2}_{D}}\int d^Dx \sqrt{-g}\left[R -\frac {4}{D-2}
(\partial\phi)^{2}-\frac{1}{4}e^{2d\phi} F^2\right],
\feq
where $\kappa_{D}$ is the D-dimensional gravitational coupling constant, $\phi$
is the dilaton field, $d$ is the dilaton coupling parameter, and $F$ is
the field strength of the $U(1)$ gauge field.

In the dual-frame \cite{Behrndt:1999mk} the static solution of the model has
the near-horizon form AdS$_{2} \times S^{D-2}$. The 0-brane admits an effective
description in terms of a two-dimensional dilaton gravity theory.
Following Ref.\ \cite{Youm:2000vn} we write
\eq
S = \frac{1}{2\kappa_{2}^{2} }\int d^2x \sqrt{-g}e^{\delta \phi}[R + \gamma
(\partial\phi)^{2}+\Lambda]\,.
\label{e1}
\feq
Setting $\kappa_{2}=1$ and redefining the dilaton $\Phi=e^{\delta\phi}$, Eq.\
(\ref {e1}) is cast in the form
\beq\lb{action}
S={1\over2}\int\gr\ d^2x\ \Phi\left(R+\al{(\partial\Phi)^2\over\Phi^2}
+\Lambda\right),
\eeq
where
\eq
\alpha= {\gamma\over\delta^{2}}={1\over D-2}\left[D-1 - {4\over
d^{2}}
\left({D-3\over D-2}\right)^{2}\right].
\feq
The field equations are
\eqn
& &R+\Lambda+ {\alpha\over\Phi^{2}}(\nabla \Phi)^{2}-
{2\alpha\over\Phi}\nabla^{2}\Phi=0\,, \lb{fea}\\
& &T_{\mu\nu}={\alpha\over\Phi}\left (\nabla_{\mu}\Phi\nabla_{\nu}\Phi-
{1\over 2}g_{\mu\nu} (\nabla \Phi)^{2}\right)-
\nabla_{\mu}\nabla_{\nu}\Phi+g_{\mu\nu} \nabla ^{2}\Phi-
{1\over 2}g_{\mu\nu}\Lambda\Phi=0\,.\lb{feb}
\feqn
The trace of Eq.\ (\ref{feb}) gives
\eq
\nabla^2\Phi-\Lambda\Phi=0\,,
\feq
which is independent from $\al$.

The two-dimensional dilaton gravity model, Eq.\ (\ref{action}), has been
extensively investigated in the literature \cite{Cadoni:1994rn,Lemos:1994py}.
For sake of completeness we briefly summarize the main results. If $\alpha<1$
Eq.\ (\ref{action}) admits the asymptotically AdS black hole solutions
\eq\lb{bh}
ds^2 = -(b^2r^2-A^2(br)^{2h})dt^2 + (b^2r^2-A^2 (br)^{2h})^{-1}dr^2\,,\quad
\Phi=\Phi_{0}(b r)^{1-2h}\,,
\feq
where
\eq
h={\alpha \over 2(\alpha-1)}\,, \quad b^{2}={\Lambda\over
2(1-h)(1-2h)}\,, \qquad h < {1\over 2}\,.
\feq
The integration constant $A$ in Eq.\ (\ref{bh}) is related to the black hole
mass $m_{bh}$ by the relation
\eq\lb{mass}
m_{bh}= {1\over 2} (1-2h) \Phi_{0}A^{2}b\,.
\feq
The scalar curvature is
\eq
R=-2\left(b^{2}+h(1-2h)A^{2}(br)^{2h}r^{-2}\right).
\feq
If $h\neq 0,1/2$ the metric (\ref{bh}) has a curvature singularity at
$r=0$. Since the geometry is asymptotically AdS the boundary at $r=\infty$ is
timelike.

The thermodynamical behavior of the black hole is characterized by the
power-law mass-temperature relation
\eq
m_{bh}= {1-2h\over 2(1-h)} \Phi_{0} \left[b (1-h) \right]^{2h-1}
\left( 2\pi T\right)^{2(1-h)}\,.
\label{temp}
\feq
Below we shall restrict attention to $-1\leq h\leq 1/2$. In this case we have
$m_{bh}\sim T^{s}$, $1\leq s\leq 4$. The entropy of the black hole is
\eq
S=2\pi (\Phi_{0})^{1\over 2(1-h)} \left[2 m_{bh}\over (1-2h)
b\right]^{1-2h\over 2(1-h)}\,.
\label{entropy}
\feq

The model (\ref{action}) includes two interesting special cases: $\alpha=0$
($h=0$) and $\alpha\to -\infty$ ($h=1/2$). The former is the so-called
Jackiw-Teitelboim (JT) model \cite {JT}. The spacetime has constant curvature,
i.e.\ is locally AdS$_{2}$, and the dilaton is linear. Since the JT model has
been widely investigated in the literature (see e.g.\
\cite{Cadoni:1995uf,Cadoni:1999ja}) we shall not discuss it
here. The second case deserves a brief discussion. Taking the limit $h=1/2$ in
Eq.\ (\ref {bh}), we obtain the AdS$_{2}$ spacetime with constant dilaton
\eq\lb{bh1}
ds^2 = -(b^2r^2-A^2{br})dt^2 + (b^2r^2-A^2 {br})^{-1}dr^2\,,\quad
\Phi=\Phi_{0}\,, \quad b^{2}= \Lambda/2\,.
\feq
Since the dilaton is constant we can interpret the solution (\ref {bh1}) as
the near-horizon regime of the extremal Reissner-Nordstr\"om black hole (or its
string theory generalizations). Setting $h=1/2$ in Eq.\ (\ref {mass}) and Eq.\
(\ref{entropy}) we find states of zero energy which are characterized by
constant nonzero entropy
\eq
S ={2\pi \Phi_{0}}\,.
\label{rnentropy1}
\feq
Equation (\ref{rnentropy1}) describes the typical behavior of extremal
Reissner-Nordstr\"om black holes in the near-horizon regime
\cite{Maldacena:1999uz,Cadoni:2000hg}. The $h=1/2$ model is equivalent to a
two-dimensional free CFT. In the limit $\alpha\to -\infty$
the field equations (\ref{fea}) and (\ref{feb}) become
\bea
\label {rn2}
&&R+\Lambda=0\,,\\
&&T_{\mu\nu}=\nabla_{\mu}\Phi\nabla_{\nu}\Phi-
{1\over 2}g_{\mu\nu} (\nabla \Phi)^{2}=0.\,\label{rn1}
\eea
Equation (\ref{rn1}) describes the energy-momentum tensor of a free
two-dimensional CFT of a single boson $\Phi$.
\section{Asymptotic symmetries}
If $h\not=0,1/2$ the spacetime (\ref {bh}) is not maximally symmetric; it
admits a single Killing vector $\cal{T}$ which generates translations in time.
In contrast, if $h=0,1/2$ Eq.\ (\ref {bh}) describes the maximally symmetric
AdS$_{2}$ spacetime with $SL(2,R)$ isometry group. For $h=0$ a suitable choice
of boundary conditions \cite{Cadoni:1999ja} shows that the ASG, i.e.\ the
isometry group that preserves the asymptotic form of the metric, is generated
by a Virasoro algebra. For a generic value of $h$ the discussion of the group
of asymptotic symmetries is more involved. The boundary conditions must indeed
allow both an ASG which is larger than $\cal{T}$ and finite associated charges
\cite{Brown:1986nw}.  We will see below that these requirements are fulfilled
only for $0\leq h \leq 1/2$.

From now on we shall restrict attention to $-1\leq h \leq 1/2$ and discuss $-1
\leq h \leq 0$ and $0< h \leq 1/2$ separately. In the first case suitable
boundary conditions for the metric and the dilaton are
\bea\label{d1a}
g_{tt} &=& -(br)^2 +\gamma_{tt} +{\cal O}\left(r^{2h}\right)\,,\nonumber \\
g_{tr} &=&\gamma_{tr}(t){(b r)^{-3}}+{\cal O}\left(r^{2h-3}\right)\,,\\
g_{rr} &=& (br)^{-2} +\gamma_{rr}(br)^{-4}+{\cal O}\left(r^{2h-4}\right)\,,
\nonumber \\
\Phi&=&\Phi_{0}\left[\rho (br)^{1-2h}+\gamma_{\Phi\Phi} (br)^{-1-2h}+
 {\cal O}\left(r^{-1}\right)\right]\,,\nonumber
\eea
where the fields $\gamma(t)$ describe deformations of the dilaton and of the
timelike boundary of the spacetime. Both ${\cal O}\left(1\right)$ deformations
in $g_{tt}$ and ${\cal O}\left(r^{-4}\right)$ in $g_{rr}$ dominate the
deformations that generate the black hole in Eq.\ (\ref{bh}). The ${\cal
O}\left(1\right)$ terms in the boundary conditions are essential to extending
the isometry group of the metric to an ASG generated by a Virasoro algebra.
However, their presence leads to divergent charges associated with the
generators of the symmetry.

If $0<h\leq\ha$ the boundary conditions are
\bea\label{d1}
g_{tt} &=& -(br)^2 +\tilde\gamma_{tt}(t)(br)^{2h} +{\cal O}\left(1\right)\,,
\nonumber \\
g_{tr} &=&\tilde\gamma_{tr}(t){(b r)^{2h-3}}+{\cal O}\left(r^{-3}\right)\,,\\
g_{rr} &=& (br)^{-2} +\tilde\gamma_{rr}(t){(br)^{2h-4}}+{\cal O}\left(r^{-4}\right)\,,
\nonumber \\
\Phi&=&\Phi_{0}\left[\rho (br)^{1-2h}+\tilde\gamma_{\Phi\Phi}(br)^{-1}+
{\cal O}\left(r^{-1-2h}\right)\right]\,.\nonumber
\eea
The ${\cal O}\left(1\right)$ deformations in $g_{tt}$ and the ${\cal
O}\left(r^{-4}\right)$ deformations in $g_{rr}$ are subleading with respect to
deformations that generate the black hole in Eq.\ (\ref{bh}). They become of
the same order only for $h=0$. In this case we have an ASG which is
characterized by a Virasoro algebra and finite charges.

The Killing vectors that generate the ASG are
\eqn
\xi^t &=& \epsilon(t) + \frac{\ddot{\epsilon}(t)}{2b^4r^2}
+ {\cal O}\left(r^{-4+\de}\right)\,,
\label{d2a}\\
\xi^r &=& -r\dot{\epsilon}(t) -\frac{\alpha^r(t)}{2}
(br)^{-1+\de} +
{\cal O}\left(r^{-3+\de}\right)\,,
\label{d2b}
\feqn
where $\de=0$ if $-1\leq h\leq0$ and $\de=2h$ if $0<h\leq \ha$. The function
$\alpha^r(t)$ describes diffeomorphisms of the two-dimensional gravity theory
that die off rapidly as $r$ goes to infinity (``pure'' gauge diffeomorphisms).
The leading terms of the Killing vectors (\ref{d2a}) and (\ref{d2b}) are
identical to the JT case \cite{Cadoni:1999ja}. The generators
$L_k$ of the ASG satisfy the Virasoro algebra

\eqn\label{Vira}
[L_k,L_l]=(k-l)L_{k+l}+{c\over 12}(k^3-k)\delta_{k+l,0}\,,
\feqn
where we allow for a nonvanishing central charge. We shall see below that the
ASG has a natural realization in terms of the conformal group in one dimension
(the Diff$_{1}$ group) which describes reparametrizations of either the circle
or the line, depending on the topology of the $r\to\infty$ boundary.

If $h\neq 1/2$ the black hole solutions (\ref{bh}) are characterized by a
non-constant dilaton. For consistency, the leading term in the asymptotic
expansion of the dilaton must be of the form (\ref{d1a}) or (\ref{d1}). 
In Ref.\ \cite{Cadoni:2000ah} it has been shown that for $h=0$ the ASG of 
the metric is broken by the nontrivial dilaton  and the
presence of a nonvanishing central charge in the Virasoro algebra is related
to the symmetry breaking. The boundary fields span a representation 
of  the conformal group. This conclusion holds also for negative values of $h$.
For $0<h<1/2$ the central charge vanishes identically. For $h=1/2$ the dilaton
is constant, so the boundary conditions imply that its (on-shell) Lie
derivative vanishes. Both the SL(2,R) isometry group and the ASG of the metric
are preserved.

Using suitable boundary conditions and introducing appropriate boundary fields
we could also consider $h<-1$. For instance, for $-2<h<-1$ we could introduce
the term $\Gamma_{tt} (br)^{-2}$ in the expansion of $g_{tt}$, the term
$\Gamma_{rr} (br)^{-6}$ in the expansion of $g_{rr}$, etc. However, larger
values of $|h|$ require an increasing number of boundary fields. So in this
paper we will consider only $ -1\leq h\leq 1/2$.
\subsection{Transformation laws and equations of motion for the boundary fields}
Using the boundary conditions (\ref{d1a}) and (\ref{d1}), and the Killing
vectors (\ref{d2a}) and (\ref{d2b}) we find the transformation laws for the
boundary fields. They are:
\beq\label{d6b}
\delta \rho = \epsilon\dot{\rho} - (1-2h)\dot{\epsilon}\rho,
\eeq
and
\eqn
\delta\gamma_{tt} &=& \epsilon\dot{\gamma}_{tt} +
2\dot{\epsilon}\gamma_{tt} -
                  {\stackrel\dots\epsilon\over b^2}+b\alpha^r\,,\label{t1} \\
\delta\gamma_{rr} &=& \epsilon\dot{\gamma}_{rr} +
2\dot{\epsilon}\gamma_{rr}+
                  2b\alpha^r\,,\label{t2} \\
\delta\gamma_{\Phi\Phi} &=& \epsilon\dot{\gamma}_{\Phi\Phi}
 + (1+2h)\dot{\epsilon}\gamma_{\Phi\Phi} +
 \frac{\ddot{\epsilon}\dot{\rho}}{2b^2}-{1\over 2}(1-2h)b\alpha^r \rho \,,
 \label{t3}
\feqn
\bea
\delta\tilde \gamma_{tt} &=& \epsilon\dot{\tilde \gamma}_{tt} +
(2-2h)\dot{\epsilon}\tilde\gamma_{tt}+b\alpha^r\,, \\
\delta\tilde\gamma_{rr} &=& \epsilon\dot{\tilde\gamma}_{rr} +
(2-2h)\dot{\epsilon}\tilde\gamma_{rr}+
                  (2-2h)b\alpha^r\,, \\
\delta\tilde\gamma_{\Phi\Phi} &=& \epsilon\dot{\tilde\gamma}_{\Phi\Phi}
 + \dot{\epsilon}\tilde\gamma_{\Phi\Phi}-{1\over 2}(1-2h)b\alpha^r \rho \,,
\lb{d6}
\eea
for $h\leq 0$ and $h>0$, respectively.

As was expected, the boundary fields $\gamma$ and $\tilde\gamma$ transform
according to a representation of the conformal group which is realized as
time-reparametrizations $\delta t=\epsilon(t)$ of the boundary. In
general the conformal dimensions of the boundary fields depend on the parameter
$h$. Anomalous pieces in the transformation law of the fields imply a
nonvanishing central charge in the Virasoro algebra. The boundary fields
$\gamma_{tr}$ and $\tilde \gamma_{tr}$ transform  as conformal fields as well.
However, their deformations are irrelevant (they do not contribute to the
charges and do not affect the dynamics of the boundary), 
so we have omitted their transformation laws for simplicity.

Let us now consider the dynamics of the boundary fields. In Ref.\
\cite{Cadoni:2000gm} it has been shown that for $h=0$ the two-dimensional
gravitational dynamics in the bulk induces a dynamics of the fields on the
boundary. This result holds also for generic values of $h<0$.  At leading
order the boundary fields satisfy the equations
\bea
&&{\ddot\r\over b^2}=(1-2h)\r\gtt-(1-2h)^2\r\gxx-2(1-4h)\gff\,,\lb{eqbond1a}\\
&&{\dot\r^2\over b^2\r}+(1-2h)^2\r\gxx+4(1-2h)\gff=0\,,\lb{eqbond1b}\\
&&(1-2h)\dot\r\gtt+\ha(1-2h)^2\r\dot\ga_{rr}+2(1-2h)\dot\ga_{\f\f}
+4h{\dot\r\over\r}\gff=0\,,\lb{eqbond1c}
\eea
and
\bea
&&\r(\tgxx-\tgtt)+2\tgff=0\,,\lb{eqbond2a}\\
&&(1-h)\dot\r\tgtt+{1-2h\over2}\r\dot{\tilde\ga}_{rr}+(2-2h)
\dot{\tilde\ga}_{\f\f}+{4h(1-h)\over1-2h}
{\dot\r\over\r}\tgff=0\,,\lb{eqbond2b}
\eea
for $h<0$ and $h>0$, respectively. Note that Eq.\ (\ref{eqbond1a}) follows from
Eqs.\ (\ref{eqbond1b}) and (\ref{eqbond1c}). The equations above determine
the dynamics on the boundary, which will be investigated in Sect.\
\ref{bondy}.
\subsection{Determination of the central charge of the Virasoro algebra}
To evaluate the central charge of the Virasoro algebra that generates the ASG
we turn to the Hamiltonian formalism \cite{Cadoni:1999ja}. With
the parametrization
\beq
\ds-N^2dt^2+\s^2(dr+N^rdt)^2,
\eeq
the Hamiltonian of the theory reads
\beq\lb{ham}
H=\int dr(N\Ht+N^r\Hx).
\eeq
As usual, $N$ and $N^r$ act as Lagrange multipliers and enforce the constraints
\bea\lb{constr}
&&\Ht=-\Pe\Ps+\s^\mo\Phi''-\s^{-2}\s'\Phi'-{\La\over2}\s\Phi+{\al\over2}
(\s\Phi^\mo\Ps^2-\s^\mo\Phi^\mo\Phi'^2)=0,\nn\\
&&\Hx=\Phi'\Pe-\s\Ps'=0,
\eea
where
\bea\lb{momenta}
&&\Ps=N^\mo(-\dot\Phi+N^r\Phi'),\nn\\
&&\Pe=N^\mo(-\dot\s+(N^r\s)')+\al\s\Phi^\mo\Ps,
\eea
are the momenta conjugate to $\s$ and $\Phi$, respectively. Here a prime
denotes derivative with respect to $r$.

For non-compact spacelike surfaces a boundary term $J$ must be added to the
Hamiltonian (\ref{ham}) to obtain well-defined variational 
derivatives \cite{Regge}.
Although the above requirement fixes only the variation $\de J$, with a
suitable choice of asymptotic boundary conditions $\de J$ can be written as a
total variation of a functional $J$ of the canonical fields on the boundary. In
our case the boundary reduces to a point and the variation of $J$ is
\bea\lb{deltaj}
\de J=-\lim_{r\to\infty}&&\left[N\left(\si^\mo\de\Phi'-\si^{-2}\Phi'\de\si-
\al\si^\mo\Phi^\mo\Phi'\de\Phi\right)\right.\nn\\
&&-\left. N'(\si^\mo\de\Phi)  +N^r(\Pe\de\Phi-\si\de\Ps)\right]\,.
\eea
Let us now consider the symmetries associated with the Killing vectors $\xi$.
These are generated by the phase space functionals \cite{Teitel}
\beq\lb{h1}
H[\xi]=\int dr \big(\xo\Ht+\xp\Hx\big)+J[\xi]\,,
\eeq
where $\xo=N\xi^t$ and $\xp=\xi^r+N^r\xi^t$.
In general, the Poisson  algebra of $H[\xi]$ yields a projective
representation of the asymptotic symmetry algebra,
\beq\lb{h2}
\{H[\xi],H[\chi]\}=H[[\xi,\chi]] +K(\xi,\chi),
\eeq
where $K(\xi,\chi)$ is a central charge.

The simplest way to calculate the central charge is to evaluate the Poisson
brackets (\ref{h2}) using the explicit expression for (\ref{h1}) obtained
above. A straightforward calculation gives
\bea\lb{h3}
K([\xi,\chi])=\lim_{r\to\infty}-({\xo}'\oo-
{\oo}'\xo)\si^\mo\Ps+({\xo}'\op-{\oo}'\xp)\si^\mo\Phi'\nonumber\\
-(\xo\op-\oo\xp)\left[{\La\over2}\si\Phi+\Ps\Pe-{\al\over2}\Big(\si\Phi^\mo\Ps^2+
\si^\mo\Phi\Phi'^2\Big)\right].
\eea
For the two-dimensional AdS space, however, the boundary at infinity is a
point, so the functional derivatives which appear in the Poisson brackets are
well-defined only for pure gauge transformations, for which $J[\xi]$ vanishes.
This problem can be overcome by defining the time-averaged generators $\hat
H[\xi]$ and charges $\hat K[\xi]$ (see
 Ref. \cite{Cadoni:1999ja})\footnote{An analogous problem was
encountered in a slightly different context
in Ref.\ \cite{Carlip:1999cy}.}
\beq\lb{h6}
\hat H[\xi]={b\over 2\pi} \int_0^{2\pi\over b} dt\, H[\xi]\,,\qquad
\hat K[\xi]={b\over 2\pi} \int_0^{2\pi\over b} dt\, K[\xi]\,.
\eeq
Here we assume that the time coordinate $t$ is periodic with period $b$.

We can obtain the Virasoro algebra (\ref{Vira}) by expanding the parameters
$\epsilon(t)$ of the Killing vectors (\ref{d2a}) and (\ref{d2b}) in Fourier
modes:
\beq\lb{h7}
\xi(t)=\sum_k {i\over b}e^{ikbt}L_k\,,
\eeq
where $L_k$ are the generators of the Virasoro algebra. The central charge of
the algebra is evaluated by substituting Eq.\ (\ref{h7}) and the 
black hole ground state solution ($m_{bh}=0$) in Eq.\ (\ref{h3}) and 
then integrating with
respect to $t$,
\beq\label{kappa}
\hat K(L_k,L_l)=\lim_{r\to\infty}-2i(1-2h)\Phi_0(br)^{-2h}k^3\de_{k+l,0}\,.
\eeq
$\hat K$ vanishes for $h>0$. In contrast, for $h<0$ we have a divergent
contribution. This is due to the infinite energy of the excitations 
on the boundary.
Using Mann's formula for the mass \cite{Mann:1993yv}, or evaluating
directly $H[L_0]$, the energy of the excitations is
\beq\lb{MM}
m={b\Phi_0\over2}[(1-2h)\tgxx+4\tgff](br)^{-2h}\,,
\eeq
which diverges for $r\to\infty$. Note that all generators $H[L_k]$ exhibit the
same divergence. Nevertheless, they span a representation of the Virasoro
algebra. Indeed, neglecting the divergences and shifting $L_0$ by a constant,
$L_0\to L_0-\Phi_0$, we obtain the Virasoro algebra in the standard form
(\ref{Vira}) with central charge
\eq\label{charge}
c=\cases{24(1-2h)\Phi_0 &$h\le 0$,\cr
c=0 &$h>0\,.$}
\feq
The central charge can also be evaluated from the algebra of the charges
$J[\xi]$ on the constraint surface $H=0$. If $h\leq 0$ from Eqs.\ (\ref{d1a}) and
(\ref{deltaj}) we have
\bea\lb{d11}
\de J[\ep]&=&\eo\left[b\ep\left((1-h)\tgtt\de\r+{(1-2h)\over2}\r\de\tgxx
+2\de\tgff+{4h\over1-2h}{\tgff\over\r}\de\r\right)\right.\nn\\
&&+\left.\dot{\ep\over b}\left(\de\dot\r+{2h\over1-2h}{\dot\r\over\r}\de\r\right)-
{\ddot\ep\over b}\de\r\right]\, (br)^{-2h}\,.
\eea
The equation above is not globally integrable in the full phase space. However,
it can be integrated in a neighborhood of the classical solution, i.e.\ near
$\r=1$ \cite{Cadoni:2000ah}. Using the equations of motion (\ref{eqbond1a})-(\ref{eqbond1c}),
 and expanding Eq.\ (\ref{d11}) around the classical solutions
$\r=1+\bar\r$ we have at the leading order in $\bar\r$
\beq \lb{d13}
J(\ep)= {\eo\over b}\left(\dot\ep\dot{\bar\r}-\ddot\ep\bar\r\right)
(br)^{-2h}+\ep m\,,
\eeq
where $m$ is the mass of the excitations (\ref{MM}).

The charge (\ref{d13}) is defined up to an additive constant that has been
fixed by setting $J(\ep=1)=m$. Since we are interested in the value of the
central charge of the Virasoro algebra (which is independent from $m$) we
consider only $m=0$, i.e.\ variations near the ground state. Equations
(\ref{h6}) imply that $J$ is defined up to a total time derivative. So we
write
\beq\lb{f1}
J(\ep)= -2{\eo\over b}(br)^{-2h}\ep\ddot{\bar\r}=\ep \Theta_{tt}\,.
\eeq
Using the transformation laws (\ref{d6b})-(\ref{t3}) with parameter $\om$, 
we obtain 
\beq\lb{f2}
\ep \de_{\om}\Theta_{tt}=\ep\left(\om\dot\Theta_{tt}+2\dot\om
\Theta_{tt}\right)+K(\ep,\om)\,,
\eeq
where
$K(\ep,\om)=2(1-2h)b^\mo\eo(br)^{-2h}(\ddot\ep\,\dot\om-\ddot\om\,\dot\ep)$, in
agreement with Eq.\ (\ref{kappa}). $\Theta_{tt}$ can be interpreted as the
one-dimensional stress-energy tensor associated with the conformal symmetry.
$\hat J(\ep)$ are the charges which generate the central extension of the
Virasoro algebra.

We may repeat the previous calculations for $h>0$. In this case $\delta J$ is
\beq\lb{d11a}
\de J[\ep]=\eo\left[b\ep\left((1-h)\gtt\de\r+{(1-2h)\over2}\r\de\gxx
+2(1-h)\de\gff+{4h(1-h)\over1-2h}{\gff\over\r}\de\r\right)\right]\,.
\eeq
We integrate again this expression near $\r=1$. Using the equations of motion
 (\ref{eqbond2a}) and
(\ref{eqbond2b}), and expanding around the classical solutions
$\r=1+\bar\r$ we obtain at the leading order in $\bar\r$
\beq
J(\ep)= \ep m,
\eeq
where $m$ is the mass of the excitations,
\beq
m={b\Phi_0\over2}[(1-2h)\gxx+4(1-h)\gff]\,.
\eeq
Since $J$ has no anomalous term, the central charge vanishes.
\subsection{Dynamics of the boundary}\label{bondy}
To get some clue on the origin of the boundary degrees of freedom when $h<0$ we
investigate the dynamics that is obeyed by the boundary fields
\cite{Cadoni:2000gm}. It is convenient to introduce new fields which are
invariant under the pure gauge diffeomorphisms parametrized by $\alpha^\mu$,
\bea
&&\beta = \ha(1-2h)\rho\gxx +2\gff\,,\nonumber \\
&&\gamma = \gtt - \frac 12 \gxx\,.
\eea
In terms of the new fields, the equations of motion
(\ref{eqbond1a})-(\ref{eqbond1c}) take the form
\bea
&&b^{-2}\ddot\rho = (1-2h)\rho\gamma - (1-4h)\beta,\label{d5a} \\
&&b^{-2}\dot\r^2+2(1-2h)\beta\r=0,\label{d5b} \\
&&(1-2h)(\dot{\rho}\gamma + \dot{\beta})+2h\be{\dot\r\over\r} = 0. \label{d5c}
\eea
Since the one-form
\eq
\zeta \equiv \left[(1-2h)\gamma+2h{\be\over\r}\right] d\rho + (1-2h)d\beta
\feq
is not exact, Eqs.\ (\ref{d5a})-(\ref{d5c}) determine a mechanical system with
one anholonomic constraint. Equation (\ref{d5b}) is a first integral of Eqs.\
(\ref{d5a}) and (\ref{d5c}).  It implies that the total energy vanishes:
\eq\label{energy}
E=T + V = {\dot{\rho}^2\over2b^2} +(1-2h) \beta\rho=0.
\feq
This condition is absent in the JT model \cite{Cadoni:2000gm},
where the total energy $E$ is not constrained.
$E$ is proportional to the mass of the excitations (\ref{MM}).

The Lagrange equations of the first kind for the fields $\varphi_i =
\{\rho,\beta,\gamma\}$ are
\eq
F_i - m_i\ddot{\varphi}_i + \lambda \zeta_i = 0\,,
\label{lagr}
\feq
where $F_i$ is the force that follows from the potential $V$ defined above,
$F_i = -\partial_i V$. $m_i$ denote the mass of the fields, $\lambda$ is a
Lagrange multiplier, and  $\zeta_i$ are the components of the one-form
$\zeta$. Setting
\eq
m_{\rho} = b^{-2}, \quad m_{\beta} = m_{\gamma} = 0\,,
\feq
the Lagrange equations (\ref{lagr}) yield Eq.\ (\ref{d5a}) and fix the Lagrange
multiplier as $\lambda=\rho$. The boundary fields $\varphi_i$ span a
representation of the full infinite dimensional group which is generated by the
Killing vectors (\ref{d2a}) and (\ref{d2b}).

The dynamical system (\ref{d5a})-(\ref{d5c}) can also be described in terms of
a harmonic oscillator coupled to an external source. Introducing the new field
$q=\rho^{1/2(1-2h)}$ of conformal dimension $-1/2$ and eliminating $\beta$ from
Eq.\ (\ref{d5b}) by means of Eq.\ (\ref{d5a}), we have
\eq
\ddot{q}  = \frac{b^2}{2}\gamma q\,. \label{dffeq}
\feq
Equation (\ref{d5b}) becomes
\eq
\frac{\dot{q}^2}{2} +{b^2\over4(1-2h)}\,\beta q^{4h}=0.
\label{dffham}
\feq
The equivalence of Eqs.\ (\ref{d5a})-(\ref{d5c}) and Eqs.\ (\ref{dffeq}) and
(\ref{dffham}) is straightforward. Equation (\ref{dffeq}) is the equation of
motion of a harmonic oscillator coupled to the external source $\gamma$. It
can be derived from the effective action
\eq
I = \int dt \left[\frac 12 \dot{q}^2 + \frac 14 b^2
    \gamma q^2\right]. \label{1daction}
\feq
The external source $\gamma$ which couples to the field $q$ is not constant.
Rather, it represents an operator of conformal dimension two. (Note that in the
calculation of $\delta I$, $\gamma$ being an external source must not  be
varied.) The action (\ref{1daction}) can be shown to be invariant (up to a
total derivative) under the conformal transformations (\ref{d6b})-(\ref{t3}). 
It is interesting to compare this result with the JT case, where the action
has an extra potential term \cite{Cadoni:2000gm} of de
Alfaro-Fubini-Furlan type \cite{dff}.

For $h>0$ Eqs.\ (\ref{eqbond2a}) and (\ref{eqbond2b}) are nondynamical, in
agreement with the absence of boundary degrees of freedom and vanishing central
charge.
\section{Conformal field theory at the horizon}
In the previous sections we have shown that the ASG of the metric (\ref{bh}) is
generated by a Virasoro algebra. Hence, the asymptotic region of the black hole
is described by a CFT. Besides $r\to\infty$, we also expect the region near the
black hole horizon to be described by a CFT. By investigating the algebra of
constraints in the presence of a boundary Carlip has shown that when the
boundary is a Killing horizon we can impose a natural set of boundary
conditions which leads to a Virasoro algebra \cite{Carlip:1999cy}. Moreover, on
a manifold with boundary the algebra of surface deformations acquires a central
term that depends uniquely on the boundary values of the dynamical fields.
Consequently, the algebra takes the form given in Eq.\ (\ref{Vira}), where the
explicit value of the central charge $c$ depends on the model under
consideration. In the case of a general dilaton gravity model the central
charge $c$ and the eigenvalue of $L_{0}$ are \cite{Carlip:1999cy}:
\eq\label{hor}
{c\over 6}=L_{0}=\Phi_{h},
\feq
where $\Phi_{h}$ is the value of the dilaton at the horizon $r_h$.

We conclude that the black hole geometry (\ref {bh}) has two conformal  regions
which are associated to the spacetime boundaries: $r=\infty$ and $r=r_{h}$. In
both cases the algebra of surface deformations has the form of a centrally
extended Virasoro algebra with central charge given in Eq. 
(\ref{charge}) and Eq. (\ref{hor}),  respectively. The
existence of the first conformal point ($r=\infty$) follows from the AdS
asymptotic behavior of the metric. The existence of the second conformal point
($r=r_{h}$) does not depend on the details of the solution, being simply a
consequence of the existence of a Killing horizon. Using the sigma model
formulation of dilaton gravity theories we can interpret the two conformal
regions as different coupling regimes of  the gravitational theory. In the next
three sections we will show that  $r=\infty$ and $r=r_{h}$ correspond to the
weak-coupled regime ($\Phi\to\infty$) and to the strong-coupled
($\Phi\to\Phi_{h}$) regime of the  gravitational theory, respectively.
\section{Sigma model formulation and conformal symmetries}
The conformal structure described in the previous sections
can be
traced back to the conformal invariance of the two-dimensional dilaton gravity
theory (\ref{action}).
Classically, the generic two-dimensional dilaton gravity
theory
\eq
S={1\over 2}\int d^2x\sqrt{-g}\,\left[\Phi R[g]-{d\,\ln|W(\Phi)|\over d\,\Phi}
(\nabla\Phi)^2+V(\Phi)\right]\,,\label{l1}
\label{m1}
\feq
is invariant under the generalized (conformal) Weyl transformation
\cite{Cavaglia:2000uw}
\eq
V(\Phi)\to V(\Phi)/\Omega(\Phi)\,,\quad
W(\Phi)\to\Omega(\Phi) W(\Phi)\,,\quad
g_{\mu\nu}\to\Omega(\Phi) g_{\mu\nu}\,.
\label{m2}
\feq
Here $W(\Phi)$, $V(\Phi)$, and $\Omega(\Phi)$ are arbitrary function of the
dilaton field. The classical conformal invariance is generally broken by
quantum effects. It is however (perturbatively) preserved for models with
vanishing beta function.

The conformal invariance of the theory can be made manifest by implementing the
canonical transformation $(\rho,\Phi)\to(M,\Phi)$ \cite{Cavaglia:1999xj}, where
$\rho$ is the conformal degree of freedom of the two-dimensional metric,
$g_{\mu\nu}=\rho(x)\,\eta_{\mu\nu}$, and
\eq
M=N(\Phi)-W(\Phi)(\nabla\Phi)^2\,,~~~~N(\Phi)=\int^\Phi
d\Phi'[W(\Phi')V(\Phi')]\,.
\label{m3}
\feq
The new field $M$ is invariant under Weyl and gauge transformations and is
locally conserved. Apart from a constant normalization factor, $M$ coincides
(on-shell) with the ADM mass of the system.
 Neglecting inessential surface
terms, in the new canonical chart the action (\ref{m1}) reads
\eq
S_\sigma={1\over 2}\int d^2x\,\sqrt{-g}\,{\nabla_\mu\Phi\nabla^\mu M
\over N(\Phi)-M}\,.
\label{m4}
\feq

Let us first discuss the $r=\infty$ conformal region of the black hole
geometry (\ref {bh}). Since the (coordinate-dependent) coupling constant of the 
gravitational model (\ref{action}) is $\Phi^{-1}$ from Eq.\ (\ref{bh}) 
it follows that $r=\infty$
corresponds to the weak-coupled regime of the theory. The weak-coupled regime
of the action (\ref{m4}) describes a free open string. It is convenient to
introduce the Weyl-rescaled frame\footnote{From now on we denote with
upper-case letters the quantities in the rescaled frame.}
\eq
G_{\mu\nu}= \Phi^{1-a} g_{\mu\nu}\,,
\label{m5}
\feq
where $a=(1-2h)^{-1}>0$. In this frame the action (\ref{action}) reads
\eq
S = {1\over 2}\int d^2X \sqrt{-G}
(\Phi R[G] +(1+a)\Phi^{a}\tilde\lambda^2)\,,
\label{m6}
\feq
where $\tilde\lambda^{2}=\lambda^{2}/(1+a)$ and $a=1-\alpha$. 
Equation (\ref{m4}) takes the
form
\eq
S_\sigma={1\over 2}\int d^2X\, \sqrt{-G} \, {\nabla_{\mu}\Phi
\nabla^{\mu}M\over \tilde\lambda^{2}\Phi^{a+1}-M}\,.
\label{m7}
\feq
At the tree-level in the weak-coupling expansion Eq.\ (\ref{m7}) describes a
bosonic string that propagates in a two-dimensional flat spacetime. When
$\Phi\to\infty$ we have
\eq
S_{string}={1\over 2\pi}\int d^2X\,\partial_\mu\xi^\alpha\partial^\mu
\xi_\alpha\,,
\label{m8}
\feq
where
\eq
M={1\over\sqrt{\pi}}\,(\xi^1+\xi^0)\,,\quad
{1\over a\tilde\lambda^2}\Phi^{-a}=-{1\over\sqrt{\pi}}\,(\xi^1-\xi^0)\,.
\label{m9}
\feq
Using the same arguments of Ref.\ \cite{Cadoni:2001fq} we can
work out a non-trivial relationship between the asymptotic symmetries of the
model (\ref{action}) and the conformal symmetries of the (Dirichlet) open string
(\ref{m8}). We find that the Weyl transformation (\ref{m5}) transforms the
conformal symmetries of the open string into the asymptotic symmetries of the
gravitational theory.

In the Weyl-rescaled frame the solution (\ref {bh}) reads \cite{Mi}
\eq
ds^2=-[(\tilde \lambda R)^{a+1}-\mu]dT^2+[(\tlambda
R)^{a+1}-\mu]^{-1}dR^2
\quad \Phi=\tlambda R\,,
\qquad R>0\,,
\label{m10}
\feq
where
\eq
\mu={M\over\tilde\lambda^2}\,.
\label{m11}
\feq
$M$ is related to the black hole mass by $M=2\tilde \lambda
\Phi_{0}^{a}m_{bh}$. Equation (\ref{m10}) can be obtained directly from Eq.\
(\ref{bh}) by rescaling the metric according to Eq.\ (\ref{m5}) and setting
\eq
\tilde\lambda R =\Phi_{0}(br)^{1/a}\,,\qquad
T=t\Phi_0^{-a}\,,
\label{m12}
\feq
where $b=\tilde\lambda a $ and $\mu=A^{2}\Phi_{0}^{1+a}$. The weak-coupled
asymptotic form of the metric (\ref{m10}) is obtained taking the limit
$R\to\infty$:
\eq
ds^2\approx-(\tilde\lambda R)^{a+1}dT^2+(\tilde\lambda R)^{-a-1}dR^2\,.
\label{m13}
\feq
The boundary is timelike for $a>0$ and lightlike for $-1<a<0$. In conformal
coordinates the asymptotic metric (\ref{m13}) is
\eq
ds^2\approx(a\tilde\lambda X)^{-{a+1\over a}}(-dT^2+dX^2)\,,
\label{m14}
\feq
where
\eq
a\tlambda X=(\tlambda R)^{-a}\,,\qquad X>0\,.
\label{m15}
\feq

The usual conformal symmetries of the string correspond to conformal
symmetries of the metric (\ref{m10}).
Using light-cone coordinates $U=(T+X)/2$ and
$V=(X-T)/2$ the conformal Killing vectors of the string are
\eq
\chi=\chi^U(U)\partial_U+\chi^V(V)\partial_V\,.
\label{m16}
\feq
Following Ref.\ \cite{Cadoni:2001fq} and imposing Dirichlet boundary
conditions on the string,
\eq
\partial_{T}\chi^{\mu}|_{boundary}=0\,,
\label{m17}
\feq
we have 
\eq
\chi^{U,V}={1\over
2}\left[\pm\epsilon(T)+\dot\epsilon(T)X\pm{\ddot\epsilon(T)\over 2}
X^2+{\cal O}(X^3)\right]\,.
\label{m18}
\feq
In the $(T,R)$ frame the conformal Killing vectors are
\eqn
\chi^T&=&\epsilon(T)+{1\over 2a^2\tlambda^2}(\tlambda R)^{-2a}\ddot\epsilon(T)+
{\cal O}(R^{-4a})\,,
\label{m19a}\\
\chi^R&=&-{1\over a}R\dot\epsilon(T)+{\cal O}(R^{1-2a})\,.
\label{m19b}
\feqn
Performing the change of coordinates (\ref{m12}) and fixing the pure  gauge
diffeomorphisms the Killing vectors (\ref{d2a}) and (\ref{d2b}) are recognized
to coincide with Eqs.\ (\ref{m19a}) and (\ref{m19b}). The Virasoro generators are
\eq
L_k= -\left[T^{k+1}+{k(k+1)\over 2a^2\tlambda^2}(\tlambda R)^{-2a}T^{k-1}+
{\cal O}(R^{-4a})\right]\partial_T
+\left[{1\over a}(k+1)RT^{k}+{\cal O}(R^{1-2a})\right]\partial_R\,.
\label{m20}
\feq
Using the Weyl transformation (\ref{m5}) in Eqs.\ (\ref{d1a}) and (\ref{d1}) we
obtain the boundary conditions for the Weyl-rescaled metric $G_{\mu\nu}$. With
a common notation for $h<0$ and $h>0$ we write
\eqn
G_{TT}&=&G^{(0)}_{TT}(R)+G^{(1)}_{TT}(T,R)\,,\nonumber\\
G_{RR}&=&G^{(0)}_{RR}(R)+G^{(1)}_{RR}(T,R)\,,\\
G_{TR}&=&G^{(0)}_{TR}(R)+G^{(1)}_{TR}(T,R)\,,\nonumber
\label{m21}
\feqn
where
\eq
G^{(0)}_{TT}=-(\tlambda R)^{a+1}\,,\qquad
G^{(0)}_{RR}=(\tlambda R)^{-a-1}\,,\qquad
G^{(0)}_{TR}=0\,,
\label{m22}
\feq
and
\eqn
G^{(1)}_{TT}&=&\gamma_{TT}(T)+\tilde\gamma_{TT}(T)(\tlambda R)^{-a+1}
+{\cal O}(R^{-2a})+{\cal O}(R^{-3a+1})\,,\nonumber\\
G^{(1)}_{TR}&=&\gamma_{TR}(T)(\tlambda R)^{-2a-1}+
\tilde\gamma_{TR}(T)(\tlambda R)^{-3a}+{\cal O}(R^{-4a-1})+{\cal O}(R^{-5a})\,,\\
G^{(1)}_{RR}&=&\gamma_{RR}(T)(\tlambda R)^{-2a-2}+
\tilde\gamma_{RR}(T)(\tlambda R)^{-3a-1}+{\cal O}(R^{-4a-2})+{\cal O}(R^{-5a-1})\,.
\nonumber
\label{m23}
\feqn
The asymptotic boundary conditions for the dilaton are
\eq\Phi=\Phi_{0}\left(\rho  (\tlambda R)+\gamma_{\Phi\Phi} (\tlambda R)^{-a}+
\tilde \gamma_{\Phi\Phi} (\tlambda R)^{-2a+1}+
{\cal O}\left(R^{-3a}\right)+{\cal O}\left(R^{-4a+1}\right)\right)\,.
\feq
The Killing vectors (\ref{m19a}) and (\ref{m19b}) generate conformal
transformations of the asymptotic fields. At the second order the asymptotic
metric (\ref{m10}) is
\eq
ds^2=-(\tilde \lambda R)^{a+1}\left[1-\mu(\tilde \lambda R)^{-a-1}\right]dT^2+
(\tilde \lambda R)^{-a-1}\left[1+\mu(\tilde \lambda R)^{-a-1}+
{\cal O}\left(R^{-2a-2}\right)\right]dR^2\,.
\label{m24}
\feq
Taking the Lie derivatives of the metric and of the dilaton with respect to the
Killing vectors (\ref{m19a}) and (\ref{m19b}) we find
\eq
\delta G_{\mu\nu}={a-1\over a}\dot\epsilon(T) G^{(0)}_{\mu\nu}+
{\cal O}(G^{(1)}_{\mu\nu})
\label{m25}
\feq
and
\eq
\delta\Phi =-{1\over a}\dot \epsilon\Phi\,.
\label{m26}
\feq
(The dilaton has been evaluated on-shell.) The conformal transformation
reduces to the asymptotic symmetry iff $a=1$ (AdS), i.e.\ when the Weyl
transformation (\ref{m5}) is the identity. The Weyl rescaling (\ref{m5})
transforms the conformal symmetries of the metric $G_{\mu\nu}$ into the
asymptotic symmetries of the metric $g_{\mu\nu}$. From Eq.\ (\ref{m5}) we have
\eq
\delta g_{\mu\nu}= \Phi^{a-1}\delta G_{\mu\nu}+
(a-1)\Phi^{a-2}G_{\mu\nu}\delta\Phi\,.
\label{m27}
\feq
Finally, substituting Eqs.\ (\ref{m25}) and (\ref{m26}) in Eq.\ (\ref{m27}) we
obtain
\eq
\delta_{L} g_{\mu\nu}={\cal O}\left (\Phi^{a-1}G^{(1)}_{\mu\nu}\right)\,.
\label{m28}
\feq
Recalling Eq.\ (\ref{m23}) it is straightforward to verify that Eq.\
(\ref{m28}) coincides with Eqs.\ (\ref{d1a}) and (\ref{d1}).

Let us now consider the $r=r_h$ conformal region. The horizon is defined by the
equation $\tilde \lambda^{2}\Phi_h^{a+1}-M=0$. Since on the horizon $\Phi^{-1}\sim
(\tilde\lambda/m_{bh})^{1/a+1}$ the latter does not belong to the weak-coupled
regime (unless we consider macroscopic black holes for which $m_{bh}>>
\tilde\lambda$). Moreover, the sigma-model description breaks down at the
horizon because the metric of the target space has a curvature singularity at
$\Phi=\Phi_h$. So the sigma model is strong-coupled and the action 
\ (\ref{m7}) cannot
be used to describe the $r=r_h$ conformal region.
\section{Duality symmetries}
The black hole horizon can still be described in the sigma model formalism if
we make use of duality symmetries. Two-dimensional dilaton gravity possesses a
symmetry \cite{Cadoni:1995vs} that acts on the space of (classical)
solutions; the field equations (\ref{fea}) and (\ref{feb}) are invariant under
duality transformations. The explicit form of the duality transformations
depends both on the gauge and on  the conformal frame. For instance, in the
gauge
\eq\lb{wrug}
ds^{2}= - Y(\si) dt^{2}+\Phi^{1-a}(\si)d\sigma^{2}\,,
\feq
the duality transformation for the Weyl-rescaled model (\ref{m6}) takes the
simple form
\eq\lb{du1}
Y\to \Phi^{a+1}\,,\qquad
\Phi\to Y^{1\over a+1}\,.
\feq
In this gauge the black hole solutions (\ref{m10}) read
\eqn\lb{sol1}
Y&=& {1\over 4}\left(e^{{a+1\over 2}\tilde\lambda \sigma}- \mu
e^{-{a+1\over 2}\tilde\lambda \sigma}\right)^{2}\,,\nonumber\\
\Phi&=& {1\over 2}\left(e^{{a+1\over 2}\tilde\lambda \sigma}+\mu
e^{-{a+1\over 2}\tilde\lambda \sigma}\right)^{2\over a+1}\,.
\feqn
Applying the duality transformation (\ref{du1}) to Eqs.\ (\ref{sol1}), we find
that they only change the mass parameter $\mu$ into $-\mu$.
So their overall effect is to interchange the zero of  $Y$ (the horizon of
the black hole) with the zero of  $\Phi$ (the singularity), namely the value
of the coupling at the horizon $\Phi=\Phi_{h}$ with the coupling at the
singularity $\Phi=0$.

In order
to implement the duality transformation in the sigma model action we need to
repeat the calculation for the (off-shell) mass function (\ref {m3}). It is
convenient to work in the Schwarzschild gauge
\eq\lb{schw}
ds^{2}=
- Y(r) dt^{2}+Y^{-1}(r)dr^{2}\,.
\feq
In this gauge $d\Phi/dr=\tilde \lambda$ and the mass function is
\eq\lb{mass1}
M=\tilde\lambda^{2}(\Phi^{a+1}-Y).
\feq
From Eqs.\ (\ref{schw}) and (\ref{wrug}) we have
$Y=\Phi^{a-1}(dr/d\sigma)^{2}$. Substituting this result in the duality relation
(\ref{du1}) we obtain
\eq\lb{du2}
Y=\Phi^{a-1}(dr/d\sigma)^{2}\to \Phi^{a+1},\qquad
\Phi^{a+1}\to \left( \Phi^{a-1}(dr/d\sigma)^{2}\right)=Y.
\feq
So the duality transformation changes the sign of the mass function. Since the
latter is a scalar function this statement is gauge independent. Now we can
implement the duality transformation in the sigma model action (\ref{m7}). The
dual action reads
\eq
S_{\sigma,dual}=-{1\over 2}\int d^2X\, \sqrt{-G} \, {\nabla_\mu\Phi\nabla^\mu M
\over \tilde\lambda^{2}\Phi^{a+1}+M}\,.
\label{sm5}
\feq
The horizon $\Phi=\Phi_{h}$ is mapped into $\Phi=0$. This value belongs to the
weak-coupled region of the dual sigma model. Therefore, we can expand the dual
action (\ref{sm5}) around $\Phi=0$. At leading order we obtain the free CFT
\eq
S_{\sigma,dual}={1\over 2}\int d^2X\, \sqrt{-G} \,
{\nabla_\mu\Phi\nabla^\mu \tilde M}\,,
\feq
where $\tilde M=-\ln(M/\tilde\lambda^{2})$. Classically, the black hole horizon is
described by a two-dimensional  CFT (bosonic string).
\section{Perturbative conformal points}
We have shown that at the classical level the gravitational dynamics in the
limit (a) $\Phi\to\infty$ and (b) $\Phi\to\Phi_h$ is described by free CFTs.
Moreover, as we have discussed at the end of section 2, the limit (c) $a\to
\infty$ ($h\to1/2$) also leads to a free CFT. The classical conformal
invariance of the sigma model (\ref{m7}) is generally broken at quantum level.
Conformal invariance is (perturbatively) preserved only for those values of the
coupling with vanishing beta function. Now we show that the classical conformal
invariance is preserved at one-loop level, i.e.\ that (a), (b), and (c) are
three perturbative conformal points of the sigma model.

Let us first consider (a) and (c). Changing coordinates in the target space to
the dimensionless fields $\xi^\alpha$
\eq
\Phi=a^{2}(\xi^1)^{-1/a},\qquad M =a^{2a+1}\tilde\lambda^2\xi^0\,,
\label{sm1}
\feq
the sigma-model action (\ref{m7}) is cast in the standard form
\eq
S_\sigma={1\over 2}\int d^2X\, \sqrt{-G} \, {\cal
G}_{\alpha\beta}\partial_\mu\xi^{\alpha}
\partial^\mu\xi^{\beta}\,,
\label{sm2}
\feq
where the metric of the target space is
\eq
{\cal G}_{\alpha\beta}= -{a/2\over a-\xi^{0}(\xi^{1})^{a+1\over a}}
\Omega_{\alpha\beta}\,,\qquad\Omega_{\alpha\beta}=
\left(\matrix{0&1\cr 1&0}\right)\,.
\label{sm3}
\feq
For $\xi^{1}=0$ ($\Phi=\infty$) and $a=\infty$ the sigma model describes a free
two-dimensional CFT. So we can consistently use the
perturbation theory around these vacua. The one-loop beta function is
\eq
\beta_{\alpha\beta}\equiv{1\over 2\pi}{\cal R}_{\alpha\beta}=
-{a+1\over 2\pi}{(\xi^{1})^{1/a}\over
\left (a-\xi^{0}(\xi^{1})^{(a+1)/a}\right)^{2}}\Omega_{\alpha\beta}\,,
\label{sm4}
\feq
where ${\cal R}_{\alpha\beta}$ is the Ricci tensor of the target space metric. The
beta function is always negative and vanishes for $\xi^{1}=0$ and
$a=\infty$. Hence, $\xi^{1}=0$ and $a=\infty$ are perturbative (stable)
conformal points of the dilaton gravity theory.

Let us now consider the behavior of the sigma model at the horizon. Using the
duality symmetries of the model the near-horizon behavior can be described by
the dual action (\ref{sm5}) in the limit $\Phi\to 0$. Introducing the
dimensionless fields
\eq
M=\tilde\lambda^{2}e^{-\xi^{0}}\,, \qquad \Phi=\xi^{1}\,,
\label{sm6}
\feq
the sigma model action (\ref{sm5}) is cast in the standard form (\ref{sm2})
with target space metric
\eq
{\cal G}_{\alpha\beta}={1/2\over 1+e^{\xi^{0}}(\xi^{1})^{a+1}}
\Omega_{\alpha\beta}.
\label{sm7}
\feq
For $\xi^{1}=0$ the sigma model describes a two-dimensional free conformal field
theory and we can consistently use perturbation theory near $\Phi=0$. The
one-loop beta function is
\eq
\beta_{\alpha\beta}={a+1\over 2\pi} {e^{\xi^{0}}(\xi^{1})^{a}\over
\left (1+e^{\xi^{0}}(\xi^{1})^{a+1}\right)^{2}}\Omega_{\alpha\beta}.
\label{sm8}
\feq
The beta function is now positive and vanishes for $\xi^{1}=0$.

The dilaton gravity model has three perturbative conformal points where it
behaves as a free two-dimensional conformal field theory. Point (a) corresponds
to the weak-coupled regime. Since the beta function is negative this point is
stable. The free conformal theory, CFT$_{\infty}$, describes the gravitational
model near the asymptotic region. Point (b) corresponds to the horizon of the
black hole. In this case the beta function is positive and the point is
perturbatively unstable. The two-dimensional free conformal theory, CFT$_h$,
describes the gravitational model near the horizon of the black hole. Point (c)
is of different nature, since it corresponds to the limit  $a\to \infty$; in
this case the model becomes {\em identically} equivalent to a free CFT. Both
the existence and structure of the conformal points have important consequences
on the derivation of the thermodynamical properties of the black hole. We will
discuss this point in the next section.
\section{Microscopic entropy of the black hole}
Though we have identified three distinct perturbative conformal points and
their corresponding CFTs, a complete description of the 0-brane dynamics is
still missing. We are dealing with a strong-coupled system which cannot be
solved exactly at the quantum level. Since we do not know how to deal with the
regions between the conformal points, the behavior of the system near these
points is not sufficient to fully characterize the properties of the theory.
Consequently, it is not clear whether the CFTs can be used to explain the
(semiclassical) thermodynamical properties of the black hole.

As far as the algebra CFT$_{\infty}$ is concerned we can reasonably identify
the eigenvalue $L_{0}^{\infty}$ of $L_{0}$ with the energy $E$ of the
gravitational configuration,  $L_{0}^{\infty}={E/ b}$. The energy $E$ is
indeed the eigenvalue of the Killing vector which generates time translations
and thus coincides with $L_{0}^{\infty}$. The two parameters that characterize
the CFT$_{\infty}$  algebra, $L_{0}^{\infty}$ and the central charge
$c^{\infty}$ (see Eq.\ (\ref{charge})), are completely determined by the energy
and the zero mode of the dilaton, i.e.\  by the observables
which are associated to the gravitational configuration. We can use the Cardy
formula \cite{Cardy:1986ie}
\eq\label{cardy}
S=2\pi\sqrt{cL _{0} \over 6}
\feq
to compute the entropy of the gravitational configuration as a function of the
density of states of CFT$_{\infty}$. Substituting Eq.\ (\ref{charge}) in Eq.\
(\ref{cardy}) we obtain
\eq\label{Scardy}
S=\displaystyle \cases{4\pi \sqrt{{(1-2h)E\Phi_{0}\over
b}} &$h\le 0$,\cr\cr
0 &$h>0$.\cr}
\feq
Setting $E=m_{bh}$ we should recover the thermodynamical entropy of the black
hole. However, for generic nonzero values of $h$ Eq.\ (\ref{Scardy}) does not
coincide with Eq.\ (\ref{entropy})\footnote{For $h=0$ (the JT model) Eq.\
(\ref{Scardy}) leads to a mismatch of a $\sqrt{2}$ factor between the
thermodynamical and the CFT entropy. However, the origin of the discrepancy is
known \cite{Cadoni:2000gm}.}. This result has different origins depending on
the sign of $h$. We stressed in Sect.\ 3 that for $h<0$ the deformations that
generate the Virasoro algebra do not generate the black hole. Therefore, we
cannot identify $E$ with the black hole mass. Conversely, for $h>0$ the
deformations that generate the Virasoro algebra also generate the black hole
and we can identify $E$ with $m_{bh}$. However, the deformations do not
correspond to truly dynamical degrees of freedom and the central charge
vanishes. This is no surprise, since  for $h\neq 0$ there is no obvious reason why the
dynamics of a strong-coupled system such as the black hole should be described
by a free CFT\footnote{For $h=0$ one can invoke the AdS/CFT correspondence.}.

As far as the CFT$_{h}$ algebra is concerned both $L_{0}^{h}$ and $c^{h}$ are
determined by the value of the dilaton at the horizon. Inserting Eq.\
(\ref{hor}) in Eq.\ (\ref{cardy}) we recover the thermodynamical entropy of the
black hole (\ref{entropy}) \cite{Carlip:1999cy}. Although the density  of
states of CFT$_{h}$ explains correctly the entropy of the black hole, both
$L_{0}^{h}$ and $c^{h}$ depend on the Hawking temperature of the horizon
$T_{h}$ (see Eq.\ (\ref{hor})). Therefore, we cannot identify $T_{h}$ with the
temperature $T_{CFT}$ of CFT$_{h}$; the energy-temperature relation for a
general CFT is $E\propto c T_{CFT}^{2}/b$. From this relation and from Eqs.\
(\ref{cardy}) and (\ref{hor}) it follows that $T_{CFT}$ is determined by the
curvature of the AdS space, $T_{CFT}\propto b$. Though we expect that the
thermodynamical properties of the black hole are described by a CFT at the
horizon, the relation between the CFT and the parameters of the black hole is
far from being trivial. 

The conclusions above have a counterpart in the sigma model formulation. The
asymptotic region represents a perturbative stable conformal fixed point. In
contrast, the horizon lies deep in the strong-coupled region of the sigma model
and its CFT description is obtained through a duality transformation whose
physical meaning is not completely clear. 

Let us conclude this section by briefly discussing the $a\to\infty$ regime. In
this case the dilaton is constant and the solution (\ref{bh1}) can be described
by a CFT everywhere. The metric (\ref{bh1}) possesses a Killing horizon, we
can use Eq.\ (\ref{hor}) to compute $L_{0}$ and $c$. Since $\Phi=\Phi_{0}$ we
obtain $c/6=L_{0}=\Phi_{0}$. Substituting the latter in the Cardy formula we
recover the thermodynamical entropy (\ref{rnentropy1}). It should be noticed
that if we used the $r\to\infty$ boundary calculation of Sect.\ 3 to compute
the central charge we would obtain $c=0$, i.e.\ $S=0$. This is consistent with
the fact that there are no dynamical degrees of freedom on the boundary of
AdS$_{2}$. The degrees of freedom which are responsible for the entropy are not
localized on the boundary of AdS$_{2}$ but on the Killing horizon of the
solution (\ref{bh1}).
\section{Conclusions}
In this paper we have investigated dilatonic 0-branes in the near-horizon
approximation. In the dual frame the solutions have the AdS$_{2} \times
S^{D-2}$ form. Using different approaches such as the canonical realization of
asymptotic symmetries as deformation algebra, the sigma model, and dualities,
we have found that the dynamics is characterized by three conformal points both
at classical and at (one-loop) quantum levels. The ensuing CFTs have been
identified by the fields which describe the boundary deformations  and/or by
the degrees of freedom of the sigma model. We have calculated the central
charges of the Virasoro algebra.

The use of the CFTs to describe the thermodynamics of black holes (in
particular their entropy) seems problematic. Though the CFT$_{\infty}$
description is natural from the black hole point of view (the black hole mass
is identified with $L_{0}$ and the central charge is a function of $\Phi_{0}$),
nevertheless it does not reproduce the black hole entropy for generic
couplings. Technically, this may be explained either by the absence of
dynamical degrees of freedom on the (AdS) boundary of the spacetime ($h>0$) or
by the impossibility of relating them to the degrees of freedom of the black
hole ($h<0$). Only in the JT model, where the spacetime is AdS$_{2}$ and there
is no curvature singularity, CFT$_\infty$ reproduces the black hole
thermodynamical relations. In this case we have a genuine realization of the
AdS$_{2}$/CFT$_{1}$ correspondence.

In contrast, the CFT at the horizon seems to give a good description of black
hole thermodynamical relations. This is  sensible because we expect the
thermal properties of black holes to be associated with the horizon. However,
the horizon lies deep in the strong-coupled region of the sigma model and in
order to describe its dynamics by a free CFT we must employ a nonperturbative
tool: the duality symmetry of the model. So from the black hole point of view
the CFT$_{h}$ description remains obscure. Both the central charge and the
eigenvalue of $L_{0}$ depend on the Hawking temperature of the horizon, whereas
we would like $c$ to depend on $\Phi_{0}$ only.  Something fundamental 
is still missing in the picture, hidden
perhaps in the full quantum dynamics of the black hole.
\section*{Acknowledgements}
We are very grateful to D. Klemm for interesting discussions and useful comments.
This work is supported in part by funds provided by the U.S.\ Department of
Energy (D.O.E.) under cooperative research agreement DE-FC02-94ER40818.

\end{document}